\documentstyle[12pt]{article}
\newcommand{\be}{\begin{equation}}
\newcommand{\ee}{\end{equation}}
\newcommand{\beas}{\begin{eqnarray*}}
\newcommand{\eeas}{\end{eqnarray*}}
\newcommand{\bea}{\begin{eqnarray}}
\newcommand{\eea}{\end{eqnarray}}

\newcommand{\nn}{\nonumber}
\newcommand{\e}{\epsilon}
\newcommand{\s}{\sigma}
\newcommand{\g}{\Sigma}
\newcommand{\ov}{\overline}
\newcommand{\un}{\underline}
\newcommand{\st}{space-time }
\begin{titlepage}
\title{
{\normalsize
\begin{flushright}
{IHEP 2001-19}\\
\end{flushright}
\vspace{12ex}}
{\bf Space-time symplectic extension}\\[4ex]
}
\author{Yu.\ F.\ Pirogov\footnote{pirogov@mx.ihep.su}\\[1ex]
{\small\it Institute for High Energy Physics,}\\[-0.5ex]
{\small\it  Protvino, RU-142284 Moscow Region, Russia}\\
{\small\it  Moscow Institute of Physics and Technology,}\\[-0.5ex]
{\small\it  Dolgoprudny, RU-141700 Moscow Region, Russia}
}
\date{}
\begin{document}
\maketitle
\vspace{2ex}

\abstract{
\noindent
It is conjectured that in the origin of \st  there lies a
symplectic rather than metric structure. 
The symplectic symmetry $Sp(2l,C)$, $l\ge 1$ instead of
the pseudo-orthogonal one $SO(1,d-1)$, $d\geq 4$ is proposed as the
space-time local structure group. A discrete sequence of the metric
space-times of the fixed dimensionalities $d=(2l)^2$ and
signatures, with $l(2l-1)$ time-like and  $l(2l+1)$ space-like
directions, defined over the set of the Hermitian second-rank
spin-tensors is considered as an alternative to the pseudo-Euclidean
extra dimensional space-times. The basic concepts of the symplectic
framework are developed in general, and the ordinary and
next-to-ordinary \st cases with $l=1,2$, respectively,  are elaborated
in more detail. In particular, the scheme provides the rationale for
the four-dimensionality and $1+3$ signature of the ordinary
space-time.
}

\thispagestyle{empty}
\end{titlepage}

\addtocounter{page}{1}
\subsection*{1\quad Introduction}

At present, the ordinary \st is  postulated to be locally the 
Minkowski space, i.e., the  pseudo-Euclidean  space of  the
dimensionality $d=4$ with the Lorentz group $SO(1,3)$ as the local
symmetry group. Nevertheless, the spinor analysis in the
Minkowski space heavily relies on the isomorphism  for the
proper noncompact groups $SO(1,3)\simeq SL(2,C)/Z_2$, as well as that
$SO(3)\simeq SU(2)/Z_2$ for their maximal compact subgroups (see,
e.g.,~\cite{streater}). Moreover, the whole relativistic field theory
in four \st dimensions can equivalently be formulated  (and in a sense
it is  even preferable) entirely in the framework of spinors of the
$SL(2,C)$ group~\cite{pen}. In this approach, to a \st point there
corresponds a Hermitian tensor of the second rank.

From this point of view, a  description of the ordinary \st by means
of the real four-vectors of the $SO(1,3)$ group, rather than by the
Hermitian tensors of  $SL(2,C)$, is nothing but the
(historically settled) tradition of the \st parametrization.
Nevertheless, right this parametrization underlies the proposed and
widely discussed \st ex\-tensions into the (locally) pseudo-Euclidean
spaces of the larger dimensionalities $d>4$ in the Kaluza-Klein
fashion (see, e.g., \cite{kk}). These
extensions assume the embedding of the local symmetry groups as
$SO(1,3)\subset SO(1,d-1)$. The pseudo-Euclidean extensions play the
crucial role in the  attempts to construct a unified theory of
all the interactions including gravity~\cite{string}.

In what follows we stick to the viewpoint that  spinors are more
fundamental objects than vectors. Thus the \st structure group
with spinors as defining representations, i.e.\ the complex symplectic
group $Sp(2,C)$, is considered to be more appropriate than the
pseudo-orthogonal group $SO(1,3)$ with vectors as defining
representations and spinors just as a kind  of artefact. In other
words, we assume that the  symplectic structure of the \st has a
deeper physical origin than the metric one though both approaches,
symplectic and pseudo-orthogonal, are formally equivalent at an
effective level  in the ordinary space-time. Then in
searching for the \st extra dimensional extensions,  a natural step
would be to look for the extensions in the symplectic framework with
the structure group $Sp(2l,C)$, $l>1$. The reason is that
the descriptions equivalent at  $l=1$ and $d=4$
can result in principally different extensions at $l>1$  and $d>4$.
This is the problem dealt with in the present paper. We develop the
basic concepts of the  general symplectic framework and elaborate in
more detail the ordinary and next-to-ordinary \st cases  with $l=1,2$,
respectively.\footnote{An early version of the study can be found
in~\cite{pir}.}

\subsection*{2\quad Structure group}

It is assumed that an underlying
physics  described effectively by a local symmetry
(structure group) constitutes the basis for  the local
properties of the space-time, i.e., for its dimensionality and
signature. Hence, to find possible
types of the \st extensions it is necessary first of all to find out
all the  structure groups isomorphic each other at $d=4$.
In addition to the well-known isomorphism of the real and complex
groups $SO(1,3)\simeq SL(2,C)/Z_2$ relevant to the ordinary
space-time, there exist the following isomorphisms (up to $Z_2$) for
the proper complex Lie groups: $SL(2,C)\simeq
SO(3,C)\simeq Sp(2,C)$ and, respectively, for their maximal compact
(real) subgroups $SU(2)\simeq SO(3)\simeq Sp(2)$. In other terms
these isomorphisms look like $A_1\simeq B_1\simeq C_1$, where the
groups considered are the first ones from the complex Cartan series:
$A_l=SL(l+1,C)$, $B_l=SO(2l+1,C)$, $C_l=Sp(2l,C)$ and similarly for
their maximal compact subgroups 
$SU(l+1)$, $SO(2l+1)$, $Sp(2l)$~(see, e.g., \cite{helgason}). Here
$l\geq 1$ means the rank of the corresponding Lie algebras. It is
equal to the half-rank of the proper noncompact Lie groups and
coincides with the rank of their maximal compact subgroups.
As the structure groups, all the groups from the above series result
in the (locally) isomorphic descriptions at $l=1$. Therefore  at
$l>1$, the extended structure groups may a priori be looked for in
each of the series with properly extended spinor space. But the
physical requirement for the existence of an invariant bilinear
product in the extended spinor space restricts the admissible types of
extension.

Namely, for all the  complex groups the complex conjugate fundamental
representations $\bar \psi$ are not equivalent to the representations
$\psi$ themselves. Besides, for all the complex series  there
is no invariant  tensor in the spinor space which would
match a spinor representation and its complex conjugate.  Hence, the
invariant bilinear product of Grassmann fields in the  form $\psi\psi$
(and $\bar \psi\bar \psi$) is
the only possible one (if any). The latter is admissible just for the
symplectic series~$C_l$. This is due to the fact that, by definition,
there exists in this case the invariant (antisymmetric) second-rank
tensor. It is to be noted, that the spinor representations of the
orthogonal groups $B_l$  are realized  by the embedding of the latter
ones into the symplectic groups $C_{2^{l-1}}$ over the
$2^l$-dimensional spinor space.
Only at $l=1,2$ there take place the isomorphisms $B_l\simeq C_l$.
The spinors being assumed to be more fundamental objects than
vectors, it is natural to consider directly the symplectic groups
which are self-sufficient for spinors,  instead of the
pseudo-orthogonal ones which inevitably should appeal to symplectic
groups for justification of the spinor representations. 

Just the existence of the alternating second-rank tensor in the
$SL(2,C)$ group is, in
essence, the {\it raison d'etre} for the  spinor analysis in four \st
dimensions being  based traditionally on this group.
The symmetry structure which  provides the alternating tensor and,  as
a result, the invariant inner product for spinors, proves to be
crucial for the whole physical theory. But this structure survives in
$Sp(2l,C)$ and is absent in $SL(l+1,C)$ at $l>1$.
This is why namely the first groups, and not the second ones, are to
be considered  as the structure groups of the extended space-time. 
Therefore, while constructing  extra dimensional space-times  we
retain symplectic structure, i.e., consider  extensions in the  series
$C_l$.  

To summarize: two alternative ways of the space-time extension can
be pictured schematically as
\bea
SO(1,3)&\simeq& Sp(2,C) \nn \\
\downarrow\ \ \ \ &&\ \ \ \ \ \downarrow\nn \\
SO(1,d-1)&\not\simeq& Sp(2l,C)\,. 
\eea
The first, commonly adopted way of extension, corresponds to the real
structure groups while the second one relies on the complex groups.
The scheme shows that the isomorphism of the real and complex groups,
valid at $d=4$ and  $l=1$, is no longer fulfilled at $d>4$ and  $l>1$.
In the first way of extension the local metric properties of
the space-time (i.e., dimensionality and signature) are put in ab
initio. In the second way, these properties should not 
be considered as the primary ones but, instead, they have to emerge as
a manifestation of the inherent symplectic structure.

\subsection*{3\quad {\boldmath Sp(2l,C)}}

Let $\psi_A$ and  $\bar\psi{}^{\bar A}\equiv (\psi_A)^*$, as
well as their respective duals $\psi^A$  and  $\bar\psi{}_{\bar
A}\equiv (\psi^A)^*$, $A$, $\bar A =1,\dots, n$ ($n=2l$) are the 
spinor representaions of $Sp(2l,C)$. It is well known
that  there exist in the spinor space  the nondegenerate  invariant
second-rank spin-tensors $\e_{AB}=-\e_{BA}$ and $\e^{AB}=-\e^{BA}$
such that $\e_{AC}\e^{CB}=\delta_A{}^B$, with $\delta_A{}^B$ being the
Kroneker symbol
(and similarly for $\e_{\bar A\bar B}\equiv(\e^{BA})^*$ and  $\e^{\bar
A\bar B}\equiv(\e_{BA})^*$). Owing to these invariant tensors the
spinor indices of the upper and lower positions  are pairwise
equivalent ($\psi_A\sim \psi^A$ and $\bar \psi_{\bar A}\sim \bar
\psi^{\bar A}$), so that there are left just two inequivalent spinor
representaions (generically, $\psi$ and $\bar\psi$). Let us call
$\psi$ and $\bar\psi$ the spinors of the first and the second kind,
respectively, and similarly for the corresponding indices $A$
and~$\bar A$.\footnote{Note that  both the type  and
position of the indices are changed under complex
conjugation, contrary the traditional definition of the dotted indices
for $SL(2,C)$ without the position change:
$(\psi _A)^* \equiv \psi^*_{\dot A}$, etc. The advantage of the
definition adopted in the  present paper is that relative to the
maximal compact subgroup $Sp(2l)$, the two types of indices $A$ and
$\bar A$ in the same position are completely indistinguishable, while
the similar $A$ and $\dot A$ would enjoy this property only after
the implicit position change for $\dot A$. }

Let us put in correspondence to an event point $P$ a second rank
spin-tensor $X_A{}^{\bar B}(P)$, which is Hermitian, i.e.,  
$X_A{}^{\bar B}= (X_B{}^{\bar A})^*\equiv\bar X^{\bar B}{}_A$, or
in other terms $X^{A\bar B}= (X_{B\bar A})^*$.  One
can define the quadratic scalar product  as
\be\label{eq:X2}
\mbox{tr\,}X\bar X\equiv X_A{}^{\bar B}\bar X_{\bar B}{}^A=
X_A{}^{\bar B}X^A{}_{\!\bar B}
= -X_{A\bar B}X^{A\bar B}= 
-X_{A\bar B}(X_{B\bar A})^*\,,
\ee
the last equality being due to the Hermiticity  of $X$. Clearly,
$\mbox{tr\,}X\bar X$ is real though not sign definite. Besides, the
spin-tensor $X\bar X$ is antisymmetric, $(X\bar X)_{AB}=-(X\bar
X)_{BA}$, and hence it can be decomposed into the
trace relative to $\e$ and a traceless part.
Under $S\in Sp(2l,C)$ one has in compact notations:
\bea
X&\to &SXS^\dagger\,,\nn\\
\bar X&\to &S^{\dagger -1}\bar X S^{-1}\,,
\eea
so that $X\bar X \to S X\bar X S^{-1}$ and $\mbox{tr\,}X\bar
X$ is invariant, indeed. In fact,  the invariant~(\ref{eq:X2}) is at
$l>1$ just the first one
in a series of independent invariants $\mbox{tr\,}(X\bar X)^k$,
$k=1,\dots,l$.  By definition,
set $\{X\}$ endowed with the structure group $Sp(2l,C)$ and the
interval between points $X_1$ and $X_1$ defined as
$\mbox{tr\,}(X_1-X_2)(\bar X_1-\bar X_2)$ constitutes 
the  symplectic space-time. The noncompact transformations from the
$Sp(2l,C)$ are counterparts of the Lorentz boosts in the ordinary
space-time, while transformations
from  the compact subgroup $Sp(2l)=Sp(2l,C)\cap SU(2l)$ correspond to
rotations. With account for translations $X_A{}^{\bar B}\to
X_A{}^{\bar B}+\Xi_A{}^{\bar B}$, where  $\Xi_A{}^{\bar B}$ is an
arbitrary constant Hermitian spin-tensor,  the whole theory in the
flat symplectic space-time  should be invariant under the
inhomogeneous symplectic group. 

Let us now fix for a while the extended boosts and
restrict ourselves  by the  extended rotations, i.e., by the maximal
compact subgroup $Sp(2l)$. Relative to  the latter, the indices of
the first and the second types are indistinguishable  in their
transformation properties ($\psi_A\sim \bar\psi_{\bar A}$), and one
can temporarily label $X_{A\bar B}$ in this case
as $X_{XY}$, where $X,Y,\dots=1,\dots,n$ generically mean spinor
indices irrespective of their kind. Hence, while
restricting by the compact subgroup one can reduce the
tensor $X_{XY}$ into two  irreducible parts, symmetric and
antisymmetric ones: $X_{XY}=\sum_{\pm}(X_{\pm})_{ XY}$, where
$(X_{\pm})_{XY}=\pm (X_{\pm})_{YX}$ have  $d_\pm=n(n\pm
1)/2$ dimensions, respectively. One gets from~(\ref{eq:X2}) the
following decomposition for the scalar product:
\be
\mbox{tr\,}X\bar X=
\sum_{\pm} (\mp 1)(X_{\pm})_{XY}[(X_{\pm})_{XY}]^*\,. 
\ee 
At $l>1$, one can further
reduce spin-tensor $X_-$ into the trace $X_-^{(0)}$ relative to $\e$
and a traceless part $X_-^{(1)}$ as
$(X_{-})_{XY}=1/\sqrt n\, X_-^{(0)}\e_{XY}+(X_-^{(1)})_{XY}$ so that
\be
\mbox{tr\,}X\bar X=X_-^{(0)2}
+(X_{-}^{(1)})_{XY}[(X_{-}^{(1)})_{XY}]^*-
(X_{+})_{XY}[(X_{+})_{XY}]^*\,. 
\ee 
As a result, the whole
extended space-time can be decomposed with respect to the rotation
group  into three irreducible subspaces of the
1, $(n-2)(n+1)/2$ and $n(n+1)/2 $ dimensions. According to their
signature and transformation properties, the first two subspaces
correspond to the time extra dimensions, the rotationally invariant
and non-invariant ones, while the third subspace corresponds to the
spatial extra dimensions. It is to be noted that the number of
components in
the extended space, and hence that in  the spatial momentum,  is equal
to the number of the noncompact transformations (boosts). Thus, for a
massive particle there exist a rest frame with zero spatial momentum.  
In the case $n=2$ there is a unique antisymmetric
tensor $(X_-)_{XY}\sim \e_{XY}$, so that the non-invariant time
subspace is empty.  

Of course, the particular decomposition of $X$ into two parts
$X_{\pm}$ is noncovariant with respect to the whole $Sp(2l,C)$ and
depends on the boosts. Nevertheless, the decomposition being valid at
any boost, the numbers  of the positive and negative
components in $\mbox{tr}X\bar X$ is invariant under the whole
$Sp(2l,C)$. In other words,  the metric signature of the symplectic
\st 
\be
\sigma_d=(\,\underbrace{+1,\dots}_{d-}\,;\underbrace{-
1, \dots}_{d_+}\,)
\ee
is invariant. Hence, at $n=2l>2$ the structure
group $Sp(2l,C)$ of the  $n$-th rank  and the  $n(n-1)$-th order, 
acting on the Hermitian second-rank spin-tensors with  $d=n^2$
components, is just a restricted subgroup of the embedding
pseudo-orthogonal group $SO(d_-,d_+)$, of the rank $n^2/2$ and the
order $n^2(n^2-1)/2$,  acting on the  pseudo-Euclidean
space of the  dimensionality $d=n^2$. What distinguishes $Sp(2l,C)$
from $SO(d_-,d_+)$, is the total set of independent  invariants
$\mbox{tr}(X\bar X)^k$, $k=1,\dots,l$. The isomorphism between
the  groups is achieved only at $l=1$, i.e., for the 
ordinary space-time $d=4$ where there is just one invariant
$\mbox{tr}X\bar X$.

It should be stressed that in the  approach under
consideration, neither the discrete set of dimensionalities,
$d=(2l)^2$, of the extended space-time, nor its  signature, nor the
existence of the rotationally invariant one-dimensional time subspace
are postulated ab initio. Rather, they are the immediate consequences
of the underlying symplectic
structure. In particular,  the latter seems to provide the unique
rationale for the four-dimensionality of the ordinary space-time, as
well as for its signature ($+---$). Namely, these properties directly
reflect  the existence of one antisymmetric and three symmetric
second-rank Hermitian spin-tensors at $l=1$.
The set of such tensors, in its turn,  is the lowest admissible
Hermitian space to accommodate the symplectic structure, 
the case $l=0$  being trivial ($d=0$). On the other
hand, right the existence of the one-dimensional time subspace
allows one to (partially) order the events 
at any fixed boosts, which serves as a basis for the causality
description. Hence, the latter may ultimately be attributed to the
underlying symplectic structure, too. At $l>1$, because of the extra
times  being mixed via boosts with the one-dimensional time,  the
causality should approximately be valid  only at small boosts.

\subsection*{4\quad {\boldmath C, P, T}}

Let us  charge double  the spinor space, i.e., for
each  $\psi_A$,  $(\psi_A)^\dagger\equiv\bar\psi^{\bar A}$ introduce
two copies $\psi_A^{\pm}$, 
$(\psi^\pm_A)^\dagger \equiv (\bar\psi^\mp)^{\bar A}$, with
$\pm$ being the ``charge'' sign.\footnote{We use here a dagger sign
for complex conjugation to show that the Grassmann fields should
undergo the change of the order in their products.} 
In analogy to the ordinary
case of $SL(2,C)$~\cite{streater}, one can define the following
discrete symmetries:
\bea\label{eq:CPT}
C&:&\psi^{\pm}_A\to \psi^{\mp}_A\,,\nn\\
P&:&\psi^{\pm}_A\to ({\psi^{\mp}_A})^\dagger\equiv
(\ov {\psi}{}^{\pm})^{\bar A}\,,\nn\\
T&:&\psi^{\pm}_A\to ({\psi^{\pm}_A})^\dagger\equiv
(\ov {\psi}{}^{\mp})^{\bar A}\,,
\eea
and hence $CPT: \psi^{\pm}_A\to \psi^{\pm}_A$ (all up to
the phase factors). Under $CPT$ invariance, only two of the discrete
operations~(\ref{eq:CPT}) are independent ones. Without charge
doubling, just one  combination
$CP\equiv T : \psi_A\to\bar\psi{}^{\bar A}$ survives. 

Now, let us introduce the Hermitian spin-tensor current
$J =J^\dagger$ as follows
\be\label{eq:J}
J_{A}{}^{\bar B}\equiv\sum_{\pm} (\pm 1)
\psi^{\pm}_{A}({\psi}{}^\pm_{B})^\dagger
=\sum_{\pm} (\pm 1)
\psi^{\pm}_{A}(\overline{\psi}{}^\mp){}^{\bar B}\,.
\ee
($\psi$'s are the Grassmann fields). Under (\ref{eq:CPT}) the current
$J_A{}^{\bar B}$ transforms as follows 
\bea\label{eq:J_AB}
C&:&J_A{}^{\bar B}\to -J_A{}^{\bar B}\,,\nn\\
P&:&J_A{}^{\bar B}\to - J_B{}^{\bar A}\,,\nn\\
T&:&J_A{}^{\bar B}\to \phantom{-} J_B{}^{\bar A}\,.
\eea
Fixing  boosts and decomposing current $J_{A\bar B}$ into  the
symmetric and antisymmetric parts,
$J_{XY}=\sum_{\pm}(J_{\pm}){}_{XY}$, one gets
from~(\ref{eq:J_AB}):
\bea
C&:&(J_{\pm})_{XY}\to -(J_{\pm})_{XY}\,,\nn\\
P&:&(J_{\pm})_{XY}\to \mp (J_{\pm}){}^{XY}\,,\nn\\
T&:&(J_{\pm})_{XY}\to \pm (J_{\pm}){}^{XY}\,.
\eea
This is in complete agreement with the signature association for the
symmetric (antisymmetric) part of the Hermitian
spin-tensor $X$ as the extended spatial (time) components.

\subsection*{5\quad {\boldmath l = 1}}

The noncompact group $Sp(2l,C)$ has $n(n+1)$ generators
$M_{AB}=(L_{AB}, K_{AB})$, $A,B=1,\dots,n$ ($n=2l$), so that
$L_{AB}=L_{BA}$ and similarly for $K_{AB}$. The  generators $L_{AB}$
are Hermitian and 
correspond to the extended rotations, whereas those $K_{AB}$ are
anti-Hermitian and correspond to the extended boosts. In the
space of the first-kind spinors $\psi_A$ these generators can be
represented as
$(\s_{AB}, i\s_{AB})$ with 
$(\s_{AB})_{CD}= 1/2 (\e_{AC}\e_{BD}+\e_{AD}\e_{BC})$, so that
$\s_{AB}=\sigma_{BA}$ and 
$(\s_{AB})_{CD}=(\s_{AB})_{DC}$, $(\s_{AB})_{C}{}^C=0$.
Similar expressions hold true in the space of the second-kind spinors
$\bar \psi_{\bar A}$. In these terms, a canonical formalism can be
developed at arbitrary $l\ge 1$.

However, in the simplest  case  $l=1$ corresponding to the ordinary
four-dimensional space-time,
there exists the isomorphism $B_1\simeq C_1$ (or $SO(3,C)\simeq
Sp(2,C)/Z_2$). Due to this property, the structure of $Sp(2,C)$ can be
brought to the form, though  equivalent mathematically, more
familiar physically.\footnote{We use here the complex group 
$SO(3,C)$ instead of the real one $SO(1,3)$ to show the close
similarity with the next case $l=2$ where there is no real structure
group. Because of the complexity
of $SO(3,C)$ one should distinguish vectors and their complex
conjugate, the latter ones being omitted for simplicity in
what follows. The same remains true for the $SO(5,C)$ case 
corresponding to $l=2$.} Namely, let us introduce for the $SO(3,C)$
group the double set of the Pauli matrices, $(\s_{i})_{A}{}^{\bar B}$
and $(\overline\s_{i})_{\bar A}{}^{B}$,
$i=1,2,3$. They should satisfy the anticommutation relations:
$\s_i\overline \s_j+\s_j\overline \s_i=2 \delta_{ij}\s_0$ and
$\ov\s_i \s_j+\ov\s_j \s_i=2 \delta_{ij}\ov\s_0$, where
$(\s_{0})_{A}{}^B\equiv \delta_A{}^B$, 
$(\ov\s_{0})_{\bar A}{}^{\bar B}\equiv \delta_{\bar A}{}^{\bar B}$ are
the Kroneker symbols and $\delta_{ij}$ is the metric tensor of
$SO(3,C)$. Among these matrices,  $\s_0$ and $\ov\s_0$  are the
only independent ones which can be chosen antisymmetric,
$(\s_0)_{AB}\equiv
\e_{AB}$ and  $(\ov\s_0)_{\bar A\bar B}\equiv \e_{\bar A\bar B}$.
On the other hand, with respect to the maximal compact subgroup
$SO(3)$, all the matrices $\s_i$, $\ov\s_i$ can be chosen
both Hermitian and symmetric as $(\s_i)_X{}^{Y}=[(\s_i)_Y{}^{X}]^*$
and $(\s_i)_{X Y}=(\s_i)_{YX}$ (and the same for $\ov\s_i$). 
The  matrices $\s_{ij}\equiv -i/2\,(\s_i\ov\s_j-\s_j\ov\s_i)$,
such that $\s_{ij}=-\s_{ji}$ and  $(\s_{ij})_{AB}=(\s_{ij})_{BA}$ (and
similarly for $(\ov\s_{ij})_{\bar A\bar B}\equiv
i/2\,(\ov\s_i\s_j-\ov\s_j\s_i)_{\bar A\bar B}$), are
not linearly independent from $\s_i$. They can be brought to the form
$(\s_{ij})_{XY}=\e_{ijk}\,(\s_k)_{XY}$, with $\e_{ijk}$ being the
Levi-Civita $SO(3,C)$ symbol. 

The matrices ($\s_{ij}, i\s_{ij}$) can be identified as the generators
$M_{ij}=(L_{ij}, K_{ij})$ of the noncompact $SO(3,C)$ group in the
space of the first-kind spinors.  Respectively, in the space of the
second-kind spinors
they are ($-\ov\s_{ij}, i\ov\s_{ij}$). The generators $L_{ij}$
of the maximal compact subgroup $SO(3)\simeq Sp(2)/Z_2$ correspond to
rotations, while those $K_{ij}$ of the noncompact
transformations describe Lorentz boosts. 
Relative to $SO(3)$ one has $\bar \s_0=\s_0$, $\bar \s_i=\s_i$ and
$\bar \s_{ij}=-\s_{ij}$. When restricted by the
maximal compact subgroup $SO(3)$, the Hermitian second-rank
spin-tensor may be decomposed in the complete set of the Hermitian
matrices ($\s_0, \s_{ij}$) with the real
coefficients: $X=1/\sqrt 2\,
(x_0\s_0+1/2\,x_{ij}\s_{ij})$, so that $\mbox{tr}X\bar
X=x_0^2-1/2\,x_{ij}^2$.
With identification $x_{ij}\equiv \e_{ijk}x_k$ one gets as
usually $\mbox{tr}X\bar X=x_0^2-x_{i}^2$.
Both the time and spatial representations being irreducible under
$SO(3)$, there takes place the usual decomposition 
$\underline 4=\underline 1\oplus \underline 3$ 
relative to the embedding $SO(3,C)\supset SO(3)$.

\subsection*{6\quad {\boldmath l = 2}}

This case corresponds to the next-to-ordinary \st symplectic
extension. Similarly to
the  case  $l=1$, there takes place the  isomorphism
$B_2\simeq C_2$,  or $SO(5,C)\simeq Sp(4,C)/Z_2$. Cases $l=1, 2$ are
the only ones when the  structure of the symplectic group  gets
simplified in terms of the complex orthogonal groups. 
The double set of Clifford  matrices $(\g_I)_A{}^{\bar B}$ and $(\ov
\g_I)_{\bar A}{}^{B}$, $I=1,\dots,5$
satisfies $\g_I\ov\g_J+\g_J\ov\g_I=2\delta_{IJ}\g_0$ 
and $\ov\g_I\g_J+\ov\g_J\g_I=2\delta_{IJ}\ov\g_0$, where
$(\g_0)_{A}{}^{B}\equiv\delta_{A}{}^{B}$, $(\ov\g_0)_{\bar A}{}^{\bar
B}\equiv \delta_{\bar A}{}^{\bar B}$ are the Kroneker symbols  and
$\delta_{IJ}$ is the
metric tensor of $SO(5,C)$.
Relative to the maximal compact subgroup $SO(5)$
they may  be chosen Hermitian, $(\g_I)_X{}^{Y}= [(\g_I)_Y{}^{X}]^*$,
but  antisymmetric $(\g_I)_{XY}=-(\g_I)_{YX}$ (and similarly for
$\ov \g_I$), like $(\g_0)_{AB}=\e_{AB}$ and $(\ov\g_0)_{\bar A\bar
B}=\e_{\bar A \bar B}$. One
can also require that $(\g_I)_X{}^X=0$. Therefore, under restriction
by $SO(5)$, six matrices $\g_0$,  $\g_I$  provide the complete
independent set for the antisymmetric matrices in the
four-dimensional spinor space. After introducing 
matrices $\g_{IJ}=-i/2 (\g_I\ov\g_J-\g_J\ov \g_I)$, so that
$\g_{IJ}=-\g_{JI}$, one gets the  symmetry condition for them:
$(\g_{IJ})_{AB}= (\g_{IJ})_{B A}$ (and similarly for
$(\ov\g_{IJ})_{ \bar  A\bar B}=i/2
(\ov\g_I\g_J-\ov\g_J \g_I)_{\bar A\bar B}$). Hence, ten  matrices
$\g_{IJ}$ (or $\ov\g_{IJ}$) make up the complete set for the symmetric
matrices in the spinor space.
Under $SO(5)$ one has $\ov \g_0= \g_0$, $\ov \g_I= \g_I$ and $\ov
\g_{IJ}= -\g_{IJ}$.  

With respect to 
$SO(5)$ the Hermitian second-rank spin-tensor $X$ may be decomposed in
the complete set of matrices $\g_0$, $\g_I$ and $\g_{IJ}$ with the
real coefficients: $X=1/2\,(x_0\g_0+x_I\g_I+ 1/2\,x_{IJ}\g_{IJ})$.   
In these terms one gets
\be
\mbox{tr}X\bar X=x_0^2+x_I^2-\frac{1}{2}x_{IJ}^2\,.
\ee
There is one more independent invariant combination of $x_0$, $x_I$
and $x_{IJ}$ stemming from the invariant $\mbox{tr}(X\bar X)^2$.
Relative to the embedding
$SO(5,C)\supset SO(5)$ one has the following  decomposition  in the
irreducible representations: 
\be\label{eq:16} 
\un {16}=\un 1\oplus \un 5\oplus\un{10}\,.
\ee
Under the discrete transformations (\ref{eq:CPT}) one gets
\bea\label{eq:PT}
P&:&x_0\to x_0,\ x_I\to x_I,\ x_{IJ}\to -x_{IJ}\,,\nn\\
T&:&x_0\to -x_0,\  x_I\to -x_I,\ x_{IJ}\to x_{IJ}\,.
\eea
This means that from the point of view of $SO(5)$, $x_I$ is the
axial vector whereas $x_{IJ}$ is the pseudo-tensor (a counterpart
of $x_{ij}=\e_{ijk}x_k$ in three spatial dimensions).
The  matrices ($\g_{IJ}$, $i\g_{IJ}$) or ($-\ov\g_{IJ}$,
$i\ov\g_{IJ}$) represent the $SO(5,C)$
generators $M_{IJ}=(L_{IJ},K_{IJ})$ in the spaces of the spinors,
respectively, of the first and the second  kinds. A particular
expression for the matrices $\g_I$, $\g_{IJ}$ in terms of $\s_0$,
$\s_i$ depends on the fashion of the embedding  
$SO(3,C)\subset SO(5,C)$. 

The rank of the algebra $C_2$ being $l=2$, an arbitrary irreducible
representation of
the noncompact group $Sp(4,C)$ is uniquely characterized by two
complex Casimir operators $I_2$ and $I_4$, respectively, of
the second and the forth order, i.e., by four real quantum numbers.
Otherwise, an
irreducible representation of $Sp(4,C)$ can be described by the mixed
spin-tensor $\Psi_{A_1\dots}^{\bar B_1 \dots}$ of a proper rank. This
spin-tensor should be traceless in any pair of the indices of the same
kind, and its symmetry in each kind of the indices should correspond
to a two-row Young tableau. In fact, there  exists the
completely antisymmetric invariant tensor of the fourth rank
$\e_{A_1A_2A_3A_4}\equiv \e_{A_1A_2}\e_{A_3A_4}
-\e_{A_1A_3}\e_{A_2A_4}+\e_{A_1A_4}\e_{A_2A_3}$ which corresponds to
the embedding $SL(4,C)\supset Sp(4,C)$ (and similarly for $\e_{\bar
A_1\bar A_2\bar A_3\bar A_4}$). By means of these invariant tensors,
three indices of the same kind with antisymmetry  are
equivalent to one index, whereas four indices with antisymmetry can be
omitted altogether. Hence, antisymmetry is
possible in no more than pairs of indices of the same kind. Therefore,
an irreducible representaion of $Sp(4,C)$  may unambiguously be
characterized by a set of four integers ($r_1,r_2;\bar r_1,\bar r_2$),
$r_1\ge r_2\ge 0$ and $\bar r_1\ge \bar r_2\ge 0$. Here $r_{1,2}$
(respectively, $\bar r_{1,2}$) are the numbers of boxes in the first
or the second rows of the proper Young tableau.  The rank of
the maximal compact subgroup $SO(5)\simeq
Sp(4)/Z_2$ (the rotation group) being  equal to $l=2$, a state in a
representaton is additionally characterized under fixed boosts  by
two additive quantum numbers, namely, the eigenvalues  of the mutually
commuting  momentum components  of $L_{IJ}$ in two different planes,
say, $L_{12}$ and $L_{45}$. Note, that in
the $Sp(2,C)$ case the Young tableaux are at most one-rowed, and an
irreducible representation is characterized by a pair of integers
$(r;\bar r)$, with the complex dimensionality of the representation
being $(r+1)(\bar r+1)$. In this case, there remains just one diagonal
component of the total angular momentum, say, $L_{12}\equiv L_3$.

\subsection*{7\quad {{\boldmath l} $\rightarrow$ {\boldmath 1}
reduction}}

The ultimate of the dimensionality in the given
approach is the discrete number $l=1,2,\dots$ corresponding to the
dimensionality $n=2l$ of the spinor space. The dimensionality
$d=(2l)^2$ of the \st appears just as
a secondary quantity. In reality, the extended \st with  $l>1$ should
compactify to the ordinary one with $l=1$ by means of the symplectic
gravity. Let us restrict ourselves
by the next-to-ordinary \st case with $l=2$. 
Three generic inequivalent types of the  spinor decomposition relative
to the embedding $Sp(4,C)\supset Sp(2,C)$ are conceivable: 
({\em i})~$\un 4= \un 2\oplus \un 2$,  ({\em ii})~ $\un 4= \un 2\oplus
\ov 2$ and ({\em iii})~$\un 4= \un 2\oplus \un 1\oplus \un 1$.

\vspace{1ex}
({\em i}) Chiral spinor doubling

\be\label{eq:2+2}
\un 4= \un 2\oplus \un 2
\ee
results in the decomposition of the Hermitian second-rank spin-tensor
$\un {16}\sim \un 4 \times \ov 4$  as 
\be\label{eq:16'}
\un {16}=4\cdot \un 4\,,
\ee
i.e., in a collection of four four-vectors (more precisely, of
three vectors and one axial vector, as follows from (\ref{eq:16})
and (\ref{eq:PT})). As for matter fermions, according to
(\ref{eq:2+2}) the number of the two-component fermions
after compactification is twice that of the number of the
four-component fermions prior compactification. If a
kind of the family structure reproduces itself during the
compactification, it is necessary that 
there should be at least two copies of the fermions in the extended
\st with  at least four copies of them  in the ordinary space-time.
For phenomenological reasons, the fermions in excess of three families
should acquire rather large effective Yukawa couplings as a
manifestation of the curled-up \st background. This is not in
principle impossible
because the two-component fermions in (\ref{eq:2+2}) distinguish
extra dimensions.  Note, that the requirement
for the renormalization group consistency  of the Standard Model (SM)
disfavours the fourth heavy chiral family in the model without a
rather low cut-off~\cite{pz}. But if  due to the decomposition 
(\ref{eq:16'}) for the gauge bosons,  there  appeared the additional
moderately heavy vector bosons  with the mass comparable to that of
the heavy fermions, this constraint could in principle be evaded and
the compactification scale $\Lambda$ could be envisaged  to
be both rather moderate and high without conflict with the SM
consistency. On the other hand, the  extra time-like dimensions
violate causality and the proper compactification scale~$\Lambda$ in
the pseudo-orthogonal  case is stated to be not less
than the Planck scale~\cite{yndurain}. Nevertheless, one may hope that
the latter restriction could somehow be abandoned in the symplectic
approach due to approximate causality here. It is to be valid at small
boosts or gravitational fields, so that the compactification
scale~$\Lambda$ could possibly be admitted to be not very high. 
For this reason, the given compactification scenario could still
survive at any $\Lambda$.

\vspace{1ex}
({\em ii}) Vector-like spinor doubling

\be\label{eq:2bar2} 
\un 4= \un 2\oplus \ov 2
\ee
results in the decomposition
\be
\un {16}=2\cdot \un 4\oplus \Big(\un 3+ \mbox{h.c.}\Big)
\oplus 2\cdot \un 1\,.
\ee
In the traditional four-vector notations one has $X\sim
(x_\mu^{(1,2)}$,
$x_{[\mu\nu]}$, $x^{(1,2)}$), $\mu,\nu=0,\dots,3$, with the tensor 
$x_{[\mu\nu]}$ being antisymmetric and all the
components $x$ being real. According to  (\ref{eq:2bar2}), after
compactification there should emerge the  pairs of the
ordinary and mirror matter fermions. For phenomenological reasons, one
should require the mirror fermions to have masses supposedly of the
order of the compactification
scale~$\Lambda$. Modulo reservations for the preceding case, this
compactification scenario could be valid at any~$\Lambda$, too.

\vspace{1ex}
({\em iii}) Spinor-scalar content

\be\label{eq:211}
\un 4= \un 2\oplus \un 1\oplus \un 1
\ee
results in 
\be
\un {16}= \un 4\oplus \Big(2\cdot\un 2+  \mbox{h.c.}\Big)
\oplus 4\cdot \un 1\,,
\ee
or in the mixed four-vector and spinor notations $X\sim (x_\mu$,
$x_A^{(1,2)}$, $x^{(1,2,3,4)}$), $A=1,2$. 
Due to~(\ref{eq:211}), there would take place the
violation of the spin-statistics connection  for matter fields in the
four-dimensional space-time if this connection  fulfilled in the
extended space-time. The scale of this violation should be determined
by the compactification scale $\Lambda$ which, in contrast with the
two preceding cases, have safely to be high enough for not to
violate causality within the experimental precision.

\subsection*{8\quad Gauge interactions}

Let $D_A{}^{\bar B}\equiv\partial_A{}^{\bar B}+igG_A{}^{\bar B}$ be
the generic covariant derivative, with $g$ being the gauge
coupling, the Hermitian spin-tensor
$G_A{}^{\bar B}$ being  the gauge fields  and $\partial_A{}^{\bar
B}\equiv \partial/
\partial X^A{}_{\bar B}$ being the ordinary derivative. Now let us
introduce the strength tensor\footnote{For simplicity, we do
not distinguish in what follows the relative column positions of the
indices of different kinds.}
\bea
&&F_{\{A_1A_2\}}^{[\bar B_1\bar B_2]}\equiv
\frac{1}{ig}D_{\{A_1}^{[\bar B_1} D_{A_2\}}^{\bar B_1]}\nn\\
&&=\frac{1}{4ig}\Big(D_{A_1}^{\bar B_1}D_{A_2}^{\bar B_2}-
D_{A_2}^{\bar B_2}D_{A_1}^{\bar B_1}+
D_{A_2}^{\bar B_1}D_{A_1}^{\bar B_2}-
D_{A_1}^{\bar B_2}D_{A_2}^{\bar B_1}\Big)
\eea
and similarly for $\ov F{}_{[A_1A_2]}^{\{\bar B_1\bar B_2\}}\equiv
(F_{\{B_2 B_1\}}^{[\bar A_2\bar A_1]})^*$, where  $\{\dots\}$ and
$[\dots]$ mean the symmetrization and antisymmetrization,
respectively.
One gets
\be
F_{\{A_1A_2\}}^{[\bar B_1\bar B_2]}=
\partial_{\{A_1}^{[\bar B_1} G_{A_2\}}^{\bar B_2]}+
ig G_{\{A_1}^{[\bar B_1} G_{A_2\}}^{\bar B_2]}
\ee
and similarly for $\ov F{}_{[A_1A_2]}^{\{\bar B_1\bar B_2\}}$. These
tensors are clearly gauge invariant. The total number of the real
components in  the tensor
$F_{\{A_1A_2\}}^{[\bar B_1\bar B_2]}$ is $2 \cdot n(n-1)/2\cdot
n(n+1)/2=n^2(n^2-1)/2$, and it exactly coincides with the number of
components of the antisymmetric second-rank tensor
$F_{[\alpha\beta]}$, $\alpha,\beta=0,1,\dots,n^2-1$, defined in the
pseudo-Euclidean space of the  $d=n^2$ dimensions. But in the
symplectic case, tensor $F$ is
reducible and splits into a trace relative to $\e$ and a traceless
part, $F=F^{(0)}+F^{(1)}$, where  
$F^{(0)}{}_{\{A_1A_2\}}^{[\bar B_1\bar B_2]}\equiv
F^{(0)}_{\{A_1A_2\}}\e^{\bar B_1\bar B_2}$ and  
$F^{(1)}{}_{\{A_1A_2\}}^{[\bar B_1\bar B_2]}\e_{\bar B_1\bar B_2}=0$
(and similarly for $\ov F_{[A_1A_2]}^{\{\bar B_1\bar B_2\}}$). Hence,
one has two independent irreducible representations with the real
dimensionalities $d_0=n(n+1)$ and $d_1=n(n-2)(n+1)^2 /2$. At
$n=4$,  one has in terms of the complex tensors of $SO(5,C)$:
$F{}^{(0)}_{[IJ]}\equiv(\g_{IJ})^{A_1A_2}F^{(0)}_{\{A_1A_2\}}$ and 
$F{}^{(1)}{}_{[IJ]}^{[\bar B_1\bar B_2]}
\equiv(\g_{IJ})^{A_1A_2}F^{(1)}{}_{\{A_1A_2\}}^{[\bar B_1\bar B_2]}$.
At $n=2$,   in terms of $SO(3,C)$ there remains only
$F{}^{(0)}_{[ij]}\equiv(\sigma_{ij})^{A_1A_2}F^{(0)}_{\{A_1A_2\}}$ or,
equivalently, $F{}^{(0)}_i\equiv 1/2\, \e_{ijk}F{}^{(0)}_{[jk]}$.

For an unbroken gauge theory with fermions, the generic gauge, fermion
and mass terms of the Lagrangian ${\cal L}= {\cal L}_G+{\cal
L}_F+{\cal L}_M$ are, respectively,
\bea\label{eq:L}
{\cal L}_G&=&\sum_{s=0,1}(c_s +i\theta_s)\,
F^{(s)}F^{(s)}+{\rm h.c.} \,,\nn\\[-0.5ex]
{\cal L}_F&=&\frac{i}{2}\sum_{\pm}(\psi ^{\pm})^\dagger\!
\stackrel{\leftrightarrow} D \psi^{\pm}\,,\nn\\
{\cal L}_M&=&\psi^+m_0\,\psi^-
+\sum_{\pm}\psi^{\pm}m_{\pm}\psi^{\pm}+{\rm h.c.}\,,
\eea
where $F^{(s)}F^{(s)}\equiv 
F^{(s)}{}_{\{A_1A_2\}}^{[\bar B_1\bar B_2]}
F^{(s)}{}^{\{A_2A_1\}}_{[\bar B_2\bar B_1]}$.
In the Lagrangian, $m_0$ is the generic Dirac mass, $m_\pm$ are
Majorana masses, $c_s$ and $\theta_s$ are the  real gauge parameters.
One of
the parameters $c_s$, supposedly
$c_0\neq 0$, can be normalized at will. 
Eq.~(\ref{eq:L}) results in the following generalization of the Dirac
equation
\be \label{eq:dir}
iD^C\!{}_{\bar B}\psi^{\pm}_C= m_0^\dagger \ov\psi{}^{\pm}_{\bar B}
+\sum_{\pm} m_\pm^\dagger\, \ov \psi{}^{\mp}_{\bar B}
\ee
and the pair of Maxwell equations ($c_0\equiv 1$ and $c_1=\theta_1=0$,
for simplicity)
\bea\label{eq:max}
(1 +i\theta_0)D^{C\bar B}F^{(0)}{}_{\{C A
\}}-\mbox{h.c.}&=&0\,,\nn\\
(1 +i\theta_0)D^{C\bar B}F^{(0)}{}_{\{C A\}}+\mbox{h.c.}
&=&2g J_{A}{}^{\bar B}\,,
\eea
with the fermion Hermitian  current $J$ given by~(\ref{eq:J}).

The tensors $F^{(s)}$, $s=1,2$ are non-Hermitian, but 
under restriction by the maximal compact subgroup $Sp(2l)$ (when there
is no distinction between the indices of different kinds) they  split
into a pair of the Hermitian ones $E^{(s)}$ and $B^{(s)}$ as follows:
$F^{(s)}=E^{(s)}+iB^{(s)}$. Here one has
$E^{(s)}{}^{[Y_1Y_2]}_{\{X_1X_2\}}
\equiv 1/2[F^{(s)}{}^{[Y_1Y_2]}_{\{X_1X_2\}}+
\big({F^{(s)}}{}^{\{X_2X_1\}}_{[Y_2Y_1]}\big)^*]$ and  
$B^{(s)}{}^{[Y_1Y_2]}_{\{X_1X_2\}}\equiv
1/2i[F^{(s)}{}^{[Y_1Y_2]}_{\{X_1X_2\}}-
\big({F^{(s)}}{}^{\{X_2X_1\}}_{[Y_2Y_1]}\big)^*]$, so that 
$E^{(s)}{}^{[Y_1Y_2]}_{\{X_1X_2\}}=
(E^{(s)}{}_{[Y_2Y_1]}^{\{X_2X_1\}})^*$ and similarly for $B^{(s)}$.
Introducing the duality transformation
$F^{(s)}\to \tilde F^{(s)}\equiv -iF^{(s)}$ with $\tilde
E^{(s)}=B^{(s)}$ and
$\tilde B^{(s)}=-E^{(s)}$, one gets 
${\cal R}e F^{(s)}F^{(s)}=E^{(s)}{}^2-B^{(s)}{}^2$ and 
${\cal I}m F^{(s)}F^{(s)}={\cal R}e \tilde
F^{(s)}F^{(s)}=2E^{(s)}B^{(s)}$. 
Though the splitting into $E^{(s)}$ and $B^{(s)}$ is noncovariant with
respect to the whole $Sp(2l,C)$, the duality transformation is
covariant. The tensors $E^{(s)}$ and $B^{(s)}$ are the counterparts of
the ordinary electric and magnetic strengths,  and $\theta_0$ is the
counterpart of the ordinary $T$-violating $\theta$-parameter for the
$n=2$ case. Thus, $\theta_1$ is an additional $T$-violating parameter
at $n>2$. Note that in the framework of  symplectic extension the
electric and
magnetic strengths stay on equal footing. This is to be compared with
the pseudo-orthogonal extension where these strengths  have  unequal
numbers of components at $d\neq 4$,
and hence there is no natural duality relation between them. The
electric-magnetic duality  of the
gauge fields (for imaginary time) play an important role  for the
study of the topolo\-gical structure of the  gauge vacuum in 
four space-time dimensions. Therefore, the similar study might be
applicable to the case of the extended symplectic space-times with
arbitrary $l>1$. 

The field equations (\ref{eq:dir}) and (\ref{eq:max}) are valid in the
flat extended \st or,
otherwise, refer to the inertial local frames. To go beyond, 
one can introduce  the Hermitian local 
frames $e\!_\alpha{}_A{}^{\bar B}(X)$,    $e\!_\alpha{}_A{}^{\bar
B}=(e\!_\alpha{}_B{}^{\bar A})^*$, with
$\alpha=0,1,\dots, n^2-1$ being 
the world vector index, the real world coordinates $x_\alpha\equiv
e\!_\alpha{}^A\!{}_{\bar B} X_A{}^{\bar B}$, as well as the
generally covariant derivative $\nabla\!_\alpha(e)$. Now, (\ref{eq:L})
can be adapted to the $d=n^2$ dimensional curved  \st equipped
with a pseudo-Riemannian structure
(the real symmetric metric $g_{\alpha\beta}(x)=
e\!_\alpha{}^A\!{}_{\bar B}
e\!_\beta{}_{A}{}^{\bar B}$),
or to the curved coordinates. In line with~\cite{uti}, one can also
supplement gauge equations by the generalized gravity equations in the
curved symplectic space-time.
But now the group of equivalence of the local frames (structure
group) is not the whole pseudo-orthogonal group $SO(d_-,d_+)$ but only
its part isomorphic to $Sp(2l,C)$.
It leaves more independent components in the local symplectic
frames compared to  the  pseudo-Rimannian frames. The  number of
components in the latter ones being equal to that in the metrics,
the symplectic gravity is not in general equivalent to the metric one. 
The curvature tensor in the symplectic case, like the
gauge one, splits additionally into irreducible parts which can
a~priori enter the gravity Lagrangian with the independent
coefficients. The ultimate reason for this 
may be that in the symplectic approach the \st is likely to be
not a fundamental entity. By this token, gravity as a generally
covariant theory of the \st distortions is to be meant just as an
effective theory. The latter admits the existence of a  number of free
parameters, the choice of which should be determined, in principle, by
the physical contents of the effective theory and should ultimately be
clarified by an underlying theory.

\subsection*{9\quad Conclusion}

The hypothesis that the symplectic structure of \st is superior to the
metric one provides, in particular, the rationale for the
four-dimensionality and $1+3$ decomposition of the ordinary
space-time. When looking for the extra dimensional \st extensions, the
hypothesis  predicts the discrete sequence of the metric space-times
of the fixed dimensionalities  and signatures. The  symplectic
extension proves to be not a~priory inconsistent  and
provides a viable alternative to the pseudo-orthogonal one. The
emerging dynamics in the extended \st is largely unorthodox and
possesses a lot of new features. The physical contents of the scheme
require further investigation. But beyond the
physical adequacy of the extra dimensional space-times, by
generalizing from the basic case $l=1$ to its counterpart for general
$l>1$, a deeper insight into the nature of the four-dimensional \st
itself may be attained. 

\vspace{0.5ex}
The author is grateful to V.V.~Kabachenko for useful discussions.

\end{document}